\documentstyle[12pt]{article}
\def\v{\begingroup\obeyspaces\u}
\def\u#1{\verb!#1!\endgroup}
\def\IQ{\v{IQ}}
\def\ISQ{\v{ISQ}}

\def\HW{{\small HERWIG}}
\def\IS{{\small ISAJET}}
\def\IW{{\small ISAWIG}}
\def\SY{{\small SUSY}}
\def\CI{{\small CIRCE}}
\def\qbar{\bar q}   
\def\Qbar{\bar Q}   
\begin{document}
\tolerance=100000
\thispagestyle{empty}
\setcounter{page}{0}
 \begin{flushright}
Cavendish-HEP-01/08\\
CERN-TH/2001-173\\
DAMTP-2001-61\\
hep-ph/0107071\\
July 2001
\end{flushright}

\begin{center}

{\Large \bf HERWIG 6.3 Release Note}\\[5mm]

{G. Corcella\\[0.5mm]
\it Department of Physics and Astronomy, University of Rochester\\[0.5mm]
E-mail: \tt{corcella@pas.rochester.edu}}\\[5mm]

{I.G.\ Knowles\\[0.5mm]
\it Department of Physics and Astronomy, University of Edinburgh\\[0.5mm]
E-mail: \tt{knowles@ph.ed.ac.uk}}\\[5mm]

{G.\ Marchesini\\[0.5mm]
\it Dipartimento di Fisica, Universit\`a di Milano-Bicocca, and I.N.F.N.,
Sezione di Milano\\[0.5mm]
E-mail: \tt{Giuseppe.Marchesini@mi.infn.it}}\\[5mm]

{S.\ Moretti\\[0.5mm]
\it Theory Division, CERN\\[0.5mm]
E-mail: \tt{stefano.moretti@cern.ch}}\\[5mm]

{K.\ Odagiri\\[0.5mm]
\it Theory Division, KEK\\[0.5mm]
E-mail: \tt{odagirik@post.kek.jp}}\\[5mm]

{P.\ Richardson\\[0.5mm]
\it Department of Applied Mathematics and Theoretical Physics and\\
Cavendish Laboratory, University of Cambridge\\[0.5mm]
E-mail: \tt{richardn@hep.phy.cam.ac.uk}}\\[5mm]

{M.H.\ Seymour\\[0.5mm]
\it Department of Physics and Astronomy,
University of Manchester\\[0.5mm]
E-mail: \tt{M.Seymour@rl.ac.uk}}\\[5mm]

{B.R.\ Webber\\[0.5mm]
\it Cavendish Laboratory, University of Cambridge\\[0.5mm]
E-mail: \tt{webber@hep.phy.cam.ac.uk}}\\[5mm]

\end{center}

\vspace*{\fill}

\begin{abstract}{\small\noindent
    A new release of the Monte Carlo program \HW\ (version 6.3) is now
    available. The main new features are new (MRST) built-in parton
    distribution functions, more SM gauge boson production processes,
    $2 \to 3$ MSSM Higgs production processes, an option to handle
    negative event weights, and an interface to the \CI\ beamstrahlung
    program.}
\end{abstract}

\vspace*{\fill}
\newpage
\tableofcontents
\setcounter{page}{1}

\section{Introduction}

The last major public version (6.2) of \HW\ was reported in detail
in \cite{AllHW}. In this note we describe the main modifications
and new features included in intermediate versions and the latest
public version, 6.3.

Please refer to \cite{AllHW} and to the present paper if
using version 6.3 of the program.

\subsection{Availability}
The new program can be obtained from the following web site:
\small\begin{quote}\tt
            http://hepwww.rl.ac.uk/theory/seymour/herwig/
\end{quote}\normalsize
    This will temporarily be mirrored at CERN for the next few weeks:
\small\begin{quote}\tt
            http://home.cern.ch/seymour/herwig/
\end{quote}\normalsize

\subsection{Increased common block size}
To take account of the increased energy and complexity of interactions at
LHC and future colliders, the default value of the parameter \v{NMXHEP},
which sets the array sizes in the standard \v{/HEPEVT/} common block,
has been increased to 4000.

\subsection{ISAWIG update}
\IW\ is an interface which allows \SY\ spectra and decay tables
generated by \IS\ \cite{ISAJET} to be read into \HW.
A new version (1.104) of \IW\ which works with \IS7.51 and includes pions
for AMSB models has been released. (Note: we have been unable to generate
SUGRA point 6 with this version of \IS.) 

The previous version of \IW, 1.103, fixed a bug in chargino and sfermion
mixing matrices. The problem was in the conversion between the convention of
\IS\ and Haber and Kane.

For fuller details see the \IW\ web site:
\small\begin{quote}\tt
http://www-thphys.physics.ox.ac.uk/users/PeterRichardson/HERWIG/isawig.html
\end{quote}\normalsize

\section{New Parton Distribution Functions}

  The default parton distributions in \HW\ were very old and did not
  include fits to any of the HERA data. In the past users could link to 
  PDFLIB in order to use more recent PDFs. However as this
  no longer seems to be maintained several new PDFs have been
  included in the new version.

\begin{table}[!h]
\begin{center}
\begin{tabular}{|c|l|}
\hline
\v{NSTRU} & Description\\
\hline
6 & Central $\alpha_S$ and gluon leading-order fit of \cite{Martin:1998np}\\
7 & Higher gluon leading-order fit of \cite{Martin:1998np}\\
8 & Average of central and higher gluon leading-order fits of \cite{Martin:1998np}\\
\hline
\end{tabular}
\end{center}
\end{table}

  It should be noted that we have only added leading-order fits because
  the evolution algorithms in \HW, in particular the backward-evolution
  algorithm for initial-state parton showering,
  are only leading-order and therefore inconsistencies could occur
  with next-to-leading-order distributions.

  The new default structure function set \v{NSTRU}=8 is the average of 
  two of the published fits \cite{Martin:1998np}, because this has been
  found \cite{Thorne} to be closer to 
  the central value of more recent next-to-leading-order fits. The
  other fits can then be used to assess the effects of varying the high-$x$
  gluon. 

\section{New Processes}
\subsection{New gauge boson production processes}

Standard Model gauge boson production gives rise to important backgrounds
for new physics searches at LHC. Two new classes of processes are included.
 
\subsubsection{IPROC=2800--2825: Gauge boson pair production}

\begin{table}[h!]
\begin{center}
\begin{tabular}{|c|l|}
\hline
\v{IPROC} & Process\\
\hline
2800 & WW production in hadron-hadron collisions\\
\hline
2810 & ZZ production in hadron-hadron collisions (including photon terms)\\
\hline
2815 & ZZ production in hadron-hadron collisions (Z only)\\
\hline
2820 & WZ production in hadron-hadron collisions (including photon terms)\\
\hline
2825 & WZ production in hadron-hadron collisions (Z only)\\
\hline
\end{tabular}
\end{center}
\end{table}

  The code already included in \HW\ for $e^+e^-\to$ WW/ZZ \cite{Gunion:1986mc} 
  was adapted for hadron-hadron collisions, including photons and the photon/Z
  interference for the resonant diagrams.

  All of these processes use a cut \v{EMMIN} (default value 20 GeV)
  on the mass of the gauge bosons produced. The cut \v{PTMIN} (default 10 GeV)
  on the transverse momentum of the bosons is also used.
  Both these cuts should not be taken to zero simultaneously if photon terms
  are included. The phase space for these processes contains a number of peaks
  and it was therefore necessary to use an adaptive multi-channel phase-space
  integration method which is described below.

  A number of new subroutines, given in Table\,\ref{tab:sub}, were added for
  these processes.

\subsubsection[IPROC=2900--2916: $ZQ\Qbar$ production]
{IPROC=2900--2916: \boldmath{$ZQ\Qbar$} production}

\begin{table}[h!]
\begin{center}
\begin{tabular}{|c|l|}
\hline
\v{IPROC} & Process\\
\hline
2900+\IQ & $gg+q\qbar\to ZQ\Qbar$ for massless $Q$ and $\Qbar$ (\IQ=1\ldots6
for $Q=d\ldots t$)\\
\hline
2910+\IQ & $gg+q\qbar\to ZQ\Qbar$, for massive $Q$ and $\Qbar$ (\IQ=1\ldots6
for $Q=d\ldots t$)\\
\hline
\end{tabular}
\end{center}
\end{table}

  The matrix elements of \cite{Kleiss:1985yh} were used for the massless
  case and an independent calculation,
  using the approach of \cite{vanEijk:1990zp},
  which was checked both for gauge invariance and against
  the massless case for the massive result.

  In both cases the decay of the Z is fully included and is selected using
  \v{MODBOS}. \v{PTMIN} controls the minimum transverse momentum of the
  outgoing quarks.
  As with gauge boson pair production it was necessary to use an optimized
  multi-channel phase-space integrator which is described below.

  A number of new subroutines, given in Table\,\ref{tab:sub}, were added for
  these processes.

\subsubsection{Optimized phase space}

  The phase space for both gauge boson pair production and $ZQ\Qbar$
  is complicated, as these processes are both treated as $2\to4$ processes.
  In order to obtain a reasonable efficiency it was necessary to adopt a
  multi-channel approach based on that described in
  \cite{Berends:1995xn,Kleiss:1994qy}.

  In each case a number of different channels are included which attempt
  to map the phase space for the different processes.
  The default weights for these different channels 
  have been chosen to optimize the efficiency for the Tevatron
  and LHC; a choice of which to use is made based on the beam energy.
  However, the choice is affected by the phase space cuts applied.
  Therefore if these are significantly altered the weights for
  the different channels need re-optimizing. 

  This is controlled by the new variable \v{OPTM} (default \v{.FALSE.}).
  If \v{OPTM}=\v{.TRUE.}, before performing the initial search for
  the maximum weight \HW\ will attempt to optimize the
  efficiency using the procedure suggested in \cite{Kleiss:1994qy}. 

  This is done by generating \v{IOPSTP} (default 5) iterations of
  \v{IOPSH} (default 100000) events.
  The choice of \v{IOPSTP} and \v{IOPSH} is a compromise between run
  time and accuracy.
  The value of \v{IOPSH} should not be significantly reduced because the
  procedure
  attempts to minimize the error on the Monte Carlo evaluation of the cross
  section and if \v{IOPSH} is small the error on the error can be significant.
  If you need to re-optimize the weights we would recommend a
  long run just to optimize the weights which can then be used in all the
  runs to generate events. 
  The new subroutine \v{HWIPHS} was added to initialize the phase space.

\subsection[New $2\to3$ MSSM Higgs production processes]{New
\boldmath{$2\to3$} MSSM Higgs production processes}

A large number of final states involving the production
of both neutral and charged Higgs bosons of the Minimal Supersymmetric
Standard Model
(MSSM) have been made available in version 6.3.  
They all proceed via $2\to3$ body hard scattering subprocesses.
They are listed below, with corresponding process numbers
(\IQ\ and \ISQ\ are as detailed in the following subsections).
Further details of their implementation can be found in \cite{SUSYpap}.

The new subroutines introduced to administer the following processes
are {\tt HWHIBQ} and {\tt HWH2BH} for
{\tt IPROC}=3500, plus {\tt HWHISQ} and {\tt HWH2SH} for 
{\tt IPROC}=3100, 3200. In addition, {\tt HWHIGQ} has been modified
to accommodate the {\tt IPROC}=3800 series. Finally, the matrix
element {\tt NME}=200, describing the $1\to 3$ body heavy-quark decays
via a virtual $H^\pm$ boson is now available and a new function,
{\tt HWDHWT}, has been introduced to this end. This can be used
to emulate e.g.\ $t\to b H^+(\to f\bar f')$ decays.

\begin{center}
\small
\begin{tabular}{|c|l|}
\hline
 \v{IPROC} &                     Process                     \\
\hline
   3100+\ISQ& $gg/q\qbar\to {\tilde q}{\tilde q}^{'*} {H^\pm}$
   (\ISQ=\v{IPROC}$-3100$ as from first table below) \\
\hline
   3200+\ISQ& $gg/q\qbar\to {\tilde q}{\tilde q}^{'*} {h,H,A}$
   (\ISQ=\v{IPROC}$-3200$ as from second table below) \\
\hline
   3500     & $b q \to b q' H^\pm$ + ch.\ conj. \\
\hline
   3710& $q\qbar \to q'\qbar' h$  \\
   3720& $q\qbar \to q'\qbar' H$  \\
\hline
   3810+\IQ& $gg+q\qbar\to Q\Qbar h$ (all $q$ flavours in $s$-channel,
                                    \IQ\ as usual for $Q$ flavour) \\
   3820+\IQ& $gg+q\qbar\to Q\Qbar H$ ('') \\
   3830+\IQ& $gg+q\qbar\to Q\Qbar A$ ('') \\
3839~~~~~~~& $gg+q\qbar\to b\bar t H^+$ + ch. conjg. (all $q$ flavours in 
$s$-channel) \\
   3840+\IQ& $gg       \to Q\Qbar h$ (\IQ\ as above) \\
   3850+\IQ& $gg       \to Q\Qbar H$ ('') \\
   3860+\IQ& $gg       \to Q\Qbar A$ ('') \\
3869~~~~~~~& $gg       \to b\bar t H^+$ + ch. conjg.  \\
   3870+\IQ& $q\qbar   \to Q\Qbar h$ (all $q$ flavours in $s$-channel,
\IQ\ as above) \\
   3880+\IQ& $q\qbar   \to Q\Qbar H$ ('') \\
   3890+\IQ& $q\qbar   \to Q\Qbar A$ ('') \\
3899~~~~~~~& $q\qbar   \to b\bar t H^+$ + ch. conjg. (all $q$ flavours in 
$s$-channel) \\
\hline
\end{tabular}
\end{center}

\subsubsection{IPROC=3110--3178: charged Higgs plus squark pair
production}\label{sqsqc}

The production of charged Higgs bosons of the MSSM in association
with squark pairs, of bottom and top flavours only, is implemented
via the 3100 series of \v{IPROC} numbers, as follows.

\begin{center}
\begin{tabular}{|c|rcl|c|}
\hline
       \v{IPROC} & partons         &$\to$& spartons         & Higgs \\
\hline
           3110  &$ gg+q\qbar $&$\to$&$ {\tilde q}_i{\tilde q}^{'*}_j$&$H^\pm$ \\
           3111  &$ gg+q\qbar $&$\to$&$ {\tilde b}_1{\tilde t}^*_1 $& $H^+$ \\
           3112  &$ gg+q\qbar $&$\to$&$ {\tilde b}_1{\tilde t}^*_2 $& $H^+$ \\
           3113  &$ gg+q\qbar $&$\to$&$ {\tilde b}_2{\tilde t}^*_1 $& $H^+$ \\
           3114  &$ gg+q\qbar $&$\to$&$ {\tilde b}_2{\tilde t}^*_2 $& $H^+$ \\
           3115  &$ gg+q\qbar $&$\to$&$ {\tilde t}_1{\tilde b}^*_1 $& $H^-$ \\
           3116  &$ gg+q\qbar $&$\to$&$ {\tilde t}_1{\tilde b}^*_2 $& $H^-$ \\
           3117  &$ gg+q\qbar $&$\to$&$ {\tilde t}_2{\tilde b}^*_1 $& $H^-$ \\
           3118  &$ gg+q\qbar $&$\to$&$ {\tilde t}_2{\tilde b}^*_2 $& $H^-$ \\
\hline
\end{tabular}
\vskip0.25cm
Add 30(60) to \v{IPROC} for $gg$($q\qbar$)-only initiated processes
\label{tab:sqch}
\end{center}
Their phenomenological relevance has been discussed in \cite{sqsqh1}.

\subsubsection{IPROC=3210--3298: neutral Higgs plus squark pair
production}\label{sqsqn}

The production of neutral Higgs bosons of the MSSM in association
with squark pairs, of bottom and top flavours only, is implemented
via the 3200 series of \v{IPROC} numbers, as follows.

\begin{center}
\begin{tabular}{|c|rcl|c|}
\hline
       \v{IPROC} & partons         &$\to$& spartons         & Higgs \\
\hline
3210(3220)[3230]&$gg+q\qbar $&$\to$&$ {\tilde q}_i{\tilde q}^*_j $& $h(H)[A]$ \\
3211(3221)[3231]&$gg+q\qbar $&$\to$&$ {\tilde b}_1{\tilde b}^*_1 $& $h(H)[A]$ \\
3212(3222)[3232]&$gg+q\qbar $&$\to$&$ {\tilde b}_1{\tilde b}^*_2 $& $h(H)[A]$ \\
3213(3223)[3233]&$gg+q\qbar $&$\to$&$ {\tilde b}_2{\tilde b}^*_1 $& $h(H)[A]$ \\
3214(3224)[3234]&$gg+q\qbar $&$\to$&$ {\tilde b}_2{\tilde b}^*_2 $& $h(H)[A]$ \\
3215(3225)[3235]&$gg+q\qbar $&$\to$&$ {\tilde t}_1{\tilde t}^*_1 $& $h(H)[A]$ \\
3216(3226)[3236]&$gg+q\qbar $&$\to$&$ {\tilde t}_1{\tilde t}^*_2 $& $h(H)[A]$ \\
3217(3227)[3237]&$gg+q\qbar $&$\to$&$ {\tilde t}_2{\tilde t}^*_1 $& $h(H)[A]$ \\
3218(3228)[3238]&$gg+q\qbar $&$\to$&$ {\tilde t}_2{\tilde t}^*_2 $& $h(H)[A]$ \\
\hline
\end{tabular}
\vskip0.25cm
Add 30(60) to \v{IPROC} for $gg$($q\qbar$)-only initiated processes
\label{tab:sqne}
\end{center}
Their phenomenological relevance has been discussed in 
\cite{sqsqh1,sqsqh2}.

\subsubsection{IPROC=3500: charged Higgs boson from
$bq$-initiated processes}\label{bq}

This process is relevant for charged Higgs scalar production at large 
$\tan\beta$ values, see \cite{SMKO}.  

\subsubsection{IPROC=3710--3720: neutral Higgs production via 
                                 weak boson fusion}

These processes are the MSSM counterparts of the SM process
of weak vector-vector fusion in hadronic collisions (\v{IPROC}=1900). They
are computed using the same subroutines and setting 
the $\Phi^0 VV$ couplings to $\sin(\beta-\alpha)$ for 
$\Phi^0=h^0$ (\v{IPROC}=3710) and to $\cos(\beta-\alpha)$ 
for $\Phi^0=H^0$ (\v{IPROC}=3720), respectively, where $V=W^\pm,Z^0$.
(There is no $A^0VV$ coupling at tree level.)

These reactions are two of the major direct production channels of neutral
CP-even Higgs bosons of the MSSM at hadron colliders, such as the 
LHC (see e.g.\ \cite{Spira}).

\subsubsection{IPROC=3811--3899: Higgs boson plus heavy quark 
                                 pair production}\label{MSSMQQHiggs}

For the case of neutral Higgs states,
\v{IPROC}=3810+\IQ\ corresponds to $h^0$ production,
\v{IPROC}=3820+\IQ\ to $H^0$ and \v{IPROC}=3830+\IQ\ to $A^0$.
(For the last case, the variable \v{PARITY} is set to $-1$.)
Note also the production of charged Higgs states, via
\v{IPROC}=3839, 3869 and 3899, in association with pairs of top-bottom 
quarks. 

In the usage of the \v{IPROC} numbers corresponding to neutral
Higgs states, when $b$-quarks are involved in $gg$-fusion modes
(\v{IPROC}(+30)=3845, 3855 or 3865),
the user should take care to avoid double-counting the chosen 
process with the corresponding $2\to1$ and $2\to 2$ cases 
of \v{IPROC}={3610}--3630 and \v{IPROC}=3410--3430
initiated by quark-antiquark annihilation, i.e.\ $b \bar b\to$ Higgs,
and (anti)quark-gluon scattering, i.e. $bg\to b$ Higgs, respectively:
see \cite{Scott}.  
Similar arguments hold for the charged Higgs states, as
the $gg$-induced process (\v{IPROC}(+30)=3869) is an alternative 
implementation of \v{IPROC}=3450 \cite{Borz}.

The associate production of neutral Higgs bosons (both CP-even and
CP-odd) of the MSSM with heavy $Q\bar Q$ pairs ($Q=b$ and $t$) is
of extreme phenomenological relevance as a discovery mode of Higgs
scalars, both at the Tevatron (Run 2) and the LHC 
(see e.g.\ \cite{Spira,TeV2}),
as is the case of the charged Higgs channel \cite{Borz,charged}.

\section{Negative Weights Option}

A Monte Carlo program generates $N_w$ weights $\{w_i\}$ such that
the estimated cross section is
\begin{equation}
\sigma =  \frac{1}{N_w}\sum_{i=1}^{N_w} w_i \equiv \overline w\;.
\end{equation}
The corresponding error depends on the width of the weight distribution:
\begin{equation}
\frac{\delta\sigma}{\sigma} = \frac{1}{\sqrt{N_w}}
\frac{\delta w_{\mbox{\scriptsize rms}}}{\overline w}\;.
\end{equation}

If only {\it positive weights} are generated, and there exists a
maximum weight $w_{\mbox{\scriptsize max}}$, then
{\it unweighted} events can be generated by `hit and miss':
$w'_i = 0$ or $1$ with probability
$P(w'_i = 1) = w_i/w_{\mbox{\scriptsize max}}$.
Then
\begin{equation}
\frac{\delta\sigma}{\sigma} = \sqrt{
\frac{w_{\mbox{\scriptsize max}}-\overline w}{N_w\overline w}}
= \sqrt{
\frac{w_{\mbox{\scriptsize max}}-\overline w}
{N_e w_{\mbox{\scriptsize max}}}}\sim\frac{1}{\sqrt{N_e}}
\end{equation}
where $N_e=N_w \overline w/w_{\mbox{\scriptsize max}}$ is the number of
{\it events} generated. The time needed (especially for detector simulation)
depends mainly on the number of events. Hence the inefficiency of
`hit and miss' is not necessarily a disaster.  This is the usual approach
adopted in Monte Carlo event generators.

{\it Negative weights} can be generated by subtraction procedures for
matrix element corrections. These are not a problem of principle but
prevent naive `hit and miss'.
To generalize `hit and miss', one can generate unweighted
events ($ w'_i = 1$) and `antievents' ($ w'_i = -1$)
with
\begin{eqnarray}\label{sign_w}
\mbox{sign}(w'_i) &=&\mbox{sign}(w_i)\;,\\ \label{prob_w}
P(w'_i = \pm 1) & =&|w_i|/|w|_{\mbox{\scriptsize max}}\;.
\end{eqnarray}
Then
\begin{equation}
\frac{\delta\sigma}{\sigma} = \sqrt{
\frac{\overline{|w|}|w|_{\mbox{\scriptsize max}}
-{\overline w}^2}{N_w{\overline w}^2}}
\sim\frac{1}{\sqrt{N_e}}\frac{\overline{|w|}}{\overline w}
\end{equation}
where $ N_e=N_w \overline{|w|}/|w|_{\mbox{\scriptsize max}}$ is
now the total number of {\it events}+{\it antievents} generated.
Again, the time needed is almost proportional to $N_e$, so this
is tolerable as long as $\overline{|w|}\sim\overline w$.
The cross section after any cuts that may be applied is
\begin{equation}\label{sigma_c}
\sigma_c = \frac{\overline{|w|}}{N_e}(N_+ - N_-)
\end{equation}
where $ N_+$ events and $ N_-$ antievents pass the cuts.

To allow for the possibility of negative weights, a new
logical parameter \v{NEGWTS} has been introduced. 
The default (\v{NEGWTS}=\v{.FALSE.}) is as before: negative weights are
forbidden. If one is detected, a non-fatal warning is issued and the
event weight is set to zero.

If \v{NEGWTS}=\v{.TRUE.}, negative weights are allowed. Statistics are
computed and printed accordingly. If unweighted events are requested
(\v{NOWGT}=\v{.TRUE.}), the initial search stores the maximum and mean
absolute weights, $|w|_{\mbox{\scriptsize max}}$ and $\overline{|w|}$.
Events and antievents are selected according to
Eqs.~(\ref{sign_w},\ref{prob_w}) and \v{EVWGT} is reset to
$\overline{|w|}\mbox{sign}(w_i)$, so that the numerator in 
Eq.~(\ref{sigma_c}) is the sum of \v{EVWGT}s for contributing
(anti)events.

\section{CIRCE Interface}

Simulation of beamstrahlung is now available, via an interface to the
\CI\ program \cite{Ohl:1997fi}.  It is implemented as a
modification of the standard bremsstrahlung implementation, so is only
available for processes for which that is available.  The produced
radiation is treated in the collinear limit, i.e.~\v{COLISR} and
the option \v{WWA} in routine \v{HWEGAM} are forcibly set \v{.TRUE.}.

For both types of distribution function, $f_{e/e}$ and $f_{\gamma/e}$,
we use a simplifying approximation: where the correct result should use
the convolution of beamstrahlung and bremsstrahlung, we actually use the
sum of the two.  For the photon distribution this is quite reasonable,
but for the electron it is perhaps more questionable.  It boils down to
the replacement:
\begin{equation}
   f_{e/e}(x) = \int_x^1 \frac{dz}z
   f_{e/e,\mathrm{brem}}\Bigl(z\Bigr)
   f_{e/e,\mathrm{beam}}\Bigl(\frac xz\Bigr)
   \to
   f_{e/e,\mathrm{brem}}(x) + f_{e/e,\mathrm{beam}}(x) - \delta(1-x).
\end{equation}
This is good to the extent that both distributions are dominated by the
$x\to1$ region.  A small utility program can be obtained from the
\HW\ web page to test this approximation.  For example, for TESLA running
at 500~GeV, the integral over $f_{e/e}(x)$ for large $x$, relevant for
threshold shapes, is accurate to 1\% for $x>0.95$ and to 10\% for
$x>0.99$, and the value of $f_{e/e}((\frac{90}{500})^2)$, relevant for
radiative return to the Z, is accurate to 2\%.

In order for the distribution to be probabilistic, the coefficient of
the delta function has to remain positive.  If it is not, \HW\ will
terminate.  This can be prevented by adjusting the \v{ZMXISR}
parameter: the default value, \v{ZMXISR}=0.999999 is too large and
\HW\ will fail, but reducing it to \v{ZMXISR}=0.99999 works fine
for all the parameter sets currently available.  It is straightforward
to check that \v{ZMXISR} values in this range do not have an
influence on any physical observables, by rerunning with it further
reduced.

The interface is controlled by five new variables, which are given in
the following table together with their default values:
\begin{center}
  \begin{tabular}{|l|l|}
    \hline
    \v{CIRCOP} & 0 \\
    \v{CIRCAC} & 2 \\
    \v{CIRCVR} & 7 \\
    \v{CIRCRV} & 9999 12 31 \\
    \v{CIRCCH} & 0 \\
    \hline
  \end{tabular}
\end{center}
\v{CIRCOP} is the main control option: \v{CIRCOP}=0 switches
off beamstrahlung and uses standard \HW; \v{CIRCOP}=1 switches
to collinear kinematics, but still leaves beamstrahlung switched off;
\v{CIRCOP}=2 uses only beamstrahlung; and \v{CIRCOP}=3 uses
both beamstrahlung and bremsstrahlung.  \v{CIRCOP}=0 and
\v{CIRCOP}=3 should therefore be regarded as `off' and `on',
respectively, with the other two options mainly for cross-checking
purposes.  The variables \v{CIRCAC}, \v{CIRCVR},
\v{CIRCRV} and \v{CIRCCH} are simply passed to \CI\ as
its input variables \v{acc}, \v{ver}, \v{rev} and
\v{chat}, as described in its documentation.  The default values
correspond to the most up-to-date revision of version~7 of the TESLA
parametrization, with minimal output.

\CI\ can be obtained from
\small\begin{quote}\tt
  http://heplix.ikp.physik.tu-darmstadt.de/nlc/beam.html
\end{quote}\normalsize

\section{Minor Changes and Bug Fixes}
\begin{itemize}

\item In the \v{INCLUDE} file, common blocks have been regrouped so that
all commons that were present in version 6.1 are unchanged.
All new variables since version 6.1 are in \v{/HWGRAV/}, \v{/HWPMRS/},
\v{/HWCIRC/}, \v{/HW6202/}, \v{/HW6203/} or \v{/HW6300/}.
From now on all common blocks will be frozen and new variables introduced
in version x.yyy will be put in a new common block \v{/HWxyyy/}. 

\item In \v{HWHDYP} lepton mass effects are corrected.

\item In \v{HWHIGS} a factor of 2 excess in the cross section for Higgs
production by gluon or quark fusion is corrected.

\item In \v{HWUDKL} safety against numerical overflows for long-lived
particles is improved.

\item In \v{HWUMBW} Breit-Wigner smearing of W and Z boson masses is enabled.

\item A work-around to overcome problems with the new DEC UNIX Fortran
compiler has been implemented.

\item In \v{IPROC}=3610--30, top quark decays now take place. Previously
tops produced in these processes were missed by the decayer and formed
top hadrons.  In addition, changes in the value of \v{IPROC} during
generation of these processes have been eliminated.

\item Gluon spin correlations have been corrected in  $gg\to A^0$
(\v{IPROC}=3630) and implemented in $gg\to$ Graviton (\v{IPROC}=4200 etc.).

\item Underlying event suppression by \v{IPROC}$\to$\v{IPROC}+10000
is now enabled in all \SY\ processes.

\item CKM matrix element bugs in W+Higgs and heavy quark+Higgs
production (\v{IPROC}/100=33,34) have been corrected.

\item The default maximum number of errors in event generation is now
set to \v{MAX(10,MAXEV/100)}.

\item  A dummy time subroutine \v{TIMEL} has been included. It should
be deleted and replaced by a system or CERN Library routine giving
\v{TRES}= CPU time remaining (seconds) if this is needed, e.g.\ to
terminate batch jobs.  The dummy returns \v{TRES}=10$^{10}$.
\end{itemize}

\begin{table}[p]
\begin{center}
\begin{tabular}{|c|l|}
\hline
Routine & Description\\
\hline
\v{CIRCEE} & Dummy \CI\ routine -- delete if using \CI \\
\v{CIRCES} & Dummy \CI\ routine -- delete if using \CI \\
\v{CIRCGG} & Dummy \CI\ routine -- delete if using \CI \\
\v{HWDHWT} & Subroutine for top decay via a virtual $H^\pm$\\
\v{HWH2BH} & Matrix element for $H^\pm$ production via $bq$-fusion \\
\v{HWH2DD} & Function   to return the $D$ function of \cite{Kleiss:1985yh} \\
\v{HWH2F1} & Subroutine to return the $F$ function of \cite{vanEijk:1990zp}
for a fixed first momentum\\
\v{HWH2F2} & Subroutine to return the $F$ function of \cite{vanEijk:1990zp}
for a fixed second momentum\\
\v{HWH2F3} & Subroutine to return the $F$ function of \cite{vanEijk:1990zp}
for all first and second momenta\\
\v{HWH2M0} & Subroutine to compute the massless matrix element for $Q\Qbar Z$\\
\v{HWH2MQ} & Subroutine to compute the massive matrix element for $Q\Qbar Z$\\
\v{HWH2PS} & Subroutine to perform the phase-space for Z+two jets\\
\v{HWH2P1} & Subroutine to select quark masses for \v{HWH2PS}\\
\v{HWH2P2} & Subroutine to select quark masses for \v{HWH2PS}\\
\v{HWH2SH} & Matrix element for squark pair plus Higgs production \\
\v{HWH2SS} & Subroutine to return the $S$ function of \cite{Kleiss:1985yh}\\
\v{HWH2T1} & Function   to return the $T_1$ function of \cite{Kleiss:1985yh}\\
\v{HWH2T2} & Function   to return the $T_2$ function of \cite{Kleiss:1985yh}\\
\v{HWH2T3} & Function   to return the $T_3$ function of \cite{Kleiss:1985yh}\\
\v{HWH2T4} & Function   to return the $T_4$ function of \cite{Kleiss:1985yh}\\
\v{HWH2T5} & Function   to return the $T_5$ function of \cite{Kleiss:1985yh}\\
\v{HWH2T6} & Function   to return the $T_6$ function of \cite{Kleiss:1985yh}\\
\v{HWH2T7} & Function   to return the $T_7$ function of \cite{Kleiss:1985yh}\\
\v{HWH2T8} & Function   to return the $T_8$ function of \cite{Kleiss:1985yh}\\
\v{HWH2T9} & Function   to return the $T_9$ function of \cite{Kleiss:1985yh}\\
\v{HWH2T0} & Function   to return the $T_{10}$ function of
\cite{Kleiss:1985yh}\\
\v{HWHDYQ} & Subroutine for $Q\Qbar Z$\\
\v{HWHGBP} & Main routine for gauge boson pair production in hadron-hadron\\
\v{HWHGBS} & Phase space for gauge boson pair production in hadron-hadron\\
\v{HWHGB1} & Selects gauge boson mass for \v{HWHGBS}\\
\v{HWHGB2} & Matrix element for WW in hadron-hadron\\
\v{HWHGB3} & Matrix element for ZZ in hadron-hadron\\
\v{HWHGB4} & Matrix element for WZ in hadron-hadron\\
\v{HWHGB5} & Selects $t$ and $u$ for \v{HWHGBS}\\
\v{HWHIBQ} & Subroutine for $H^\pm$ production via $bq$-fusion \\
\v{HWHISQ} & Subroutine for squark pair plus Higgs production \\
\v{HWHV2J} & Master subroutine for all gauge boson + two jet processes\\
\v{HWIPHS} & Subroutine to initialize the optimized phase space \\
\v{HWSMRS} & Subroutine for MRST PDFs \\
\v{TIMEL}  & Dummy time subroutine -- see Sect.~6 \\
\hline
\end{tabular} 
\end{center}
\caption{New subroutines for version 6.3.}
\label{tab:sub}
\end{table}


\begin{thebibliography}{99}

\bibitem{AllHW}
    G. Marchesini, B.R. Webber,  G. Abbiendi, I.G. Knowles, M.H. Seymour
    and L. Stanco, Comput.\ Phys.\ Commun.\ {67} (1992) 465;\\
G. Corcella, I.G. Knowles, G. Marchesini,  
S. Moretti, K. Odagiri, P. Richardson, M.H. Seymour and  B.R. Webber,
JHEP 01 (2001) 010 [hep-ph/9912396].

\bibitem{ISAJET} H.\ Baer, F.E.\ Paige, S.D.\ Protopopescu, X.\ Tata,
preprint BNL-HET-99-43, FSU-HEP-991218, UH-511-952-00, hep-ph/0001086.
 
\bibitem{Martin:1998np}
A.D.~Martin, R.G.~Roberts, W.J.~Stirling and R.S.~Thorne,
Phys.\ Lett.\ B443 (1998) 301
[hep-ph/9808371].

\bibitem{Thorne}
R.S.\ Thorne, private communication.
\bibitem{Gunion:1986mc}
J.~F.~Gunion and Z.~Kunszt,
Phys.\ Rev.\ D33 (1986) 665.

\bibitem{Kleiss:1985yh}
R.~Kleiss and W.~J.~Stirling,
Nucl.\ Phys.\ B262 (1985) 235.
\bibitem{vanEijk:1990zp}
B.~van Eijk and R.~Kleiss,
In Proceeding of the LHC Workshop, Aachen 1990, vol. 2 183-194.
\bibitem{Berends:1995xn}
F.~A.~Berends, R.~Pittau and R.~Kleiss,
Comput.\ Phys.\ Commun.\  85 (1995) 437
[hep-ph/9409326].

\bibitem{Kleiss:1994qy}
R.~Kleiss and R.~Pittau,
Comput.\ Phys.\ Commun.\  83 (1994) 141
[hep-ph/9405257].

\bibitem{SUSYpap} S.\ Moretti, K.\ Odagiri, P.\ Richardson, 
M.H.\ Seymour and B.R.\ Webber, in preparation.

\bibitem{sqsqh1} 
A. Dedes and S. Moretti,  
Eur. Phys. J. C10 (1999) 515; preprint
RAL-TR-1999-067, June 1999,
{\tt  hep-ph/9909526};\\
A. Djouadi {\it et al.}, preprint February 2000, {\tt hep-ph/0002258}.


\bibitem{sqsqh2} 
A. Dedes and S. Moretti,  
Phys. Rev. {D60} (1999) {015007};\\
A. Djouadi, J.-L. Kneur and G. Moultaka,
{Phys.~Rev.~Lett.} 80 (1998) 1830; Nucl. Phys. B569 (2000) 53;\\
G. B\'elanger, F. Boudjema and K. K. Sridhar, Nucl.Phys. B568 (2000) 3.

\bibitem{SMKO} 
S. Moretti and K. Odagiri, Phys. Rev. {D55} (1997) 5627.

\bibitem{Spira}
M. Spira, Fortsch. Phys. {46} (1998) 203.

\bibitem{Scott} D. Dicus, T. Stelzer, Z. Sullivan and  S. Willenbrock,
 Phys. Rev. D59 (1999) 094016.

\bibitem{Borz} F. Borzumati, J.-L. Kneur and N. Polonsky,
 Phys. Rev. D60 (1999) 115011;\\
S. Moretti and D.P. Roy, Phys. Lett. B470 (1999) 209.

\bibitem{TeV2} M. Carena {\it el al.}, preprint October 2000,
{\tt hep-ph/0010338}.

\bibitem{charged} 
D.J. Miller, S. Moretti, D.P. Roy and W.J. Stirling,
 Phys. Rev. D61 (2000) 055011.

\bibitem{Ohl:1997fi}
T.~Ohl,
Comput.\ Phys.\ Commun.\ 101 (1997) 269
[hep-ph/9607454].

\end{thebibliography}
\end{document}